# A supramolecular ferroelectric with two sublattices and polarization dependent conductivity

H. Mager[1], M. Litterst[1], Sophia Klubertz[1], Shyamkumar V. Haridas[2], Oleksandr Shyshov[2], M. von Delius[*,2] and M. Kemerink[*,1]

[1] Institute for Molecular Systems Engineering and Advanced Materials, Heidelberg University, Im Neuenheimer Feld 225, 69120 Heidelberg, Germany

[2] Institute of Organic Chemistry, University of Ulm, Albert-Einstein-Allee 11, 89081 Ulm, Germany

*Corresponding author e-mail: max.vondelius@uni-ulm.de; martijn.kemerink@uni-heidelberg.de

**Abstract**

The possibility to combine and finetune properties of functional molecular materials by chemical design is particularly relevant for organic ferroelectrics. In this work, we investigate a class of organic molecular materials that show long-range supramolecular organization into fibrillar bundles. In solid state, the material shows ferroelectric behavior resulting from two largely independent dipolar moieties that show up as two separate coercive fields in polarization-hysteresis and capacitance-voltage curves. Moreover, the material shows a long-range electronic conductivity that arises due to oxidation at the positive electrode, followed by electron transfer between neighboring molecules. We find that this conductivity is modulated by the direction and degree of ferroelectric polarization, which we interpret in terms of injection barrier modulation at low electric fields and a recently developed framework for asymmetric polaron hopping at high fields. With two distinct, partially independent dipolar moieties offering the possibility to use ferroelectric properties to modulate conductance, the materials presented herein are a promising basis for multifunctional materials.

**Introduction**

Ferroelectricity was first discovered in Rochelle salt by *J. Vasalek* in in 1921[1], opening up an extensive field of research. Ferroelectric materials are characterized by a bistable electric polarization that is switchable by application of an external electrical field. Furthermore, all ferroelectric materials are also piezo- and pyroelectric, making them relevant for many applications including microphones, transducers, actuators, sensors and memory chips.[2–4] While Rochelle salt contains ions of the organic tartaric acid, the following years of research and discoveries were dominated by purely inorganic materials. Barium titanate (BTO) and lead zirconium titanate (PZT) emerged as two of the leading candidates for applications, owed to their high polarization values, fast switching, high retention and stability.[5–9] For organic materials, the co-polymer poly(vinylidene fluoride-trifluoroethylene) P(VDF-TrFE), consisting of polyvinylidene fluoride (PVDF) and poly(trifluoroethylene) (PTrFE), offers the best performance metrics and has even seen commercial use.[10–16] In general, organic ferroelectrics are flexible, solution processable, potentially energy-efficient to produce and lack toxic or rare elements. In the field of wearable electronics and for medical and biological applications, organic ferroelectric materials therefore enable unprecedented functionality, courtesy of their softness and non-toxicity.[17–20] However, organic ferroelectrics still lack in performance and stability compared to their inorganic counterparts.

While extensive research is done to find new organic ferroelectric materials with optimized parameters for existing applications, another direction is the (re)search for multifunctional materials. These materials combine a variation of optical, conductive, magnetic, ferroelectric and other properties in one material.[21–24] They are, on the one hand, of considerable interest with regard to the fundamental questions regarding the origin and the possible interplay of aforementioned properties. On the other hand, they may enable entirely new applications.

In this article we present a class of supramolecular ferroelectrics that show remarkable conductive properties, which depend on the polarization state of the material. A discussion of the origin and nature of the conductivity, which does not rely on the presence of a π-system, has been reported in previous work.[25] In short, the dipolar polarization in the materials leads to a lowering of the energetic barrier at the (metal) electrode-organic interface, enabling oxidation by electron extraction at the positive electrode and consecutive electron hopping between the molecules. The supramolecular polymerization of the material occurring at elevated fields and temperatures enhances the conductivity significantly. Here, we focus on the experimental ferroelectric characterization, including experimental evidence for the existence of two different ferroelectric sub-lattices, and the interplay of ferroelectric polarization and conductivity. As it turns out, the coupling between both properties presents a challenge when trying to extract common ferroelectric parameters. We present different approaches in overcoming these hurdles and how to obtain approximate values for relevant ferroelectric parameters like the coercive field. By comparing ferroelectric behavior of molecular derivatives, we assign individual contributions, appearing as separate coercive fields, in ferroelectric measurements to the corresponding ferroic sublattice.

Conductive switching in organic ferroelectrics has been shown in semiconducting materials, i.e., materials that do contain an extended π-system, for the injection limited[26] and the bulk limited cases.[27] Casellas *et al.* even demonstrated the coexistence of both mechanisms in a single device.[21] Although similar in phenomenology and potential applications, these processes are fundamentally different from those governing ferroelectric tunnel junctions, which have the drawback of requiring extraordinary thin films.[28] In all of these examples, the ferroelectric polarization modulates the conducting properties of the charge carrying material, either by changing the energetic barriers at the (metal)electrode-organic interface or directly influencing the bulk conductivity by introducing an asymmetric potential into site energies involved in charge hopping.[27,29] In principle, this multifunctionality allows the creation of memory cells based on crossbar arrays sandwiching just one (organic) material, simplifying the architecture of memory devices.[16]

**Results and Discussion**

The investigated molecules are shown in Fig. 1a and b, and are abbreviated as **FCH-C3-A** and **FCH-E**, respectively. Full chemical names and information on the synthesis procedure are given in section 1 of the Supplementary Information (SI). **FCH-C3-A** consists of a phenyl ring attached to long alkyl chains that facilitate solubility. More importantly for this work, the molecule contains two strong dipolar units, an all-*cis* 1,2,3,4,5,6-hexafluorocyclohexane unit (6.2 Debye) and an amide group (3.7 Debye) that is connected through a three-carbon (C3) spacer. The predicted dipole moments are for single monomers, while for head-to-tail stacked units, the net dipole moment will depend on the number of units stacked and is expected to exceed the sum of its units.[30,31] The **FCH-E** reference molecule lacks the amide-C3 bridging unit of **FCH-C3-A**, but is otherwise structurally identical. Shyshov et al. showed previously that **FCH-C3-A** can undergo non-covalent polymerization in solution, leading to the formation of supramolecular double helices that are transferable to solid state by simple spin coating on a substrate.[32] We observed similar behavior in solid thin films, where field annealing at elevated temperatures results in formation of supramolecular fibers aligned in the direction of the applied external field, as shown by the AFM height image in Fig. 1e. This is reminiscent of the formation of supramolecular columns in the ferroelectric liquid-crystalline benzene-1,3,5-tricarboxamide (BTA), where the long-range order facilitates long-range dipolar alignment.[33,34] For **FCH-C3-A**, a schematic depiction of the possible orientation of the dipolar moieties in a supramolecular fiber is shown in Fig. 1f. Hence, after fiber formation, the two dipolar moieties are assumed to form spatially separated sub-stacks. The supramolecular polymerization in **FCH-C3-A** has been shown in previous work to enhance the conductivity of the material significantly.[25] Its influence on the ferroelectric properties is investigated below.

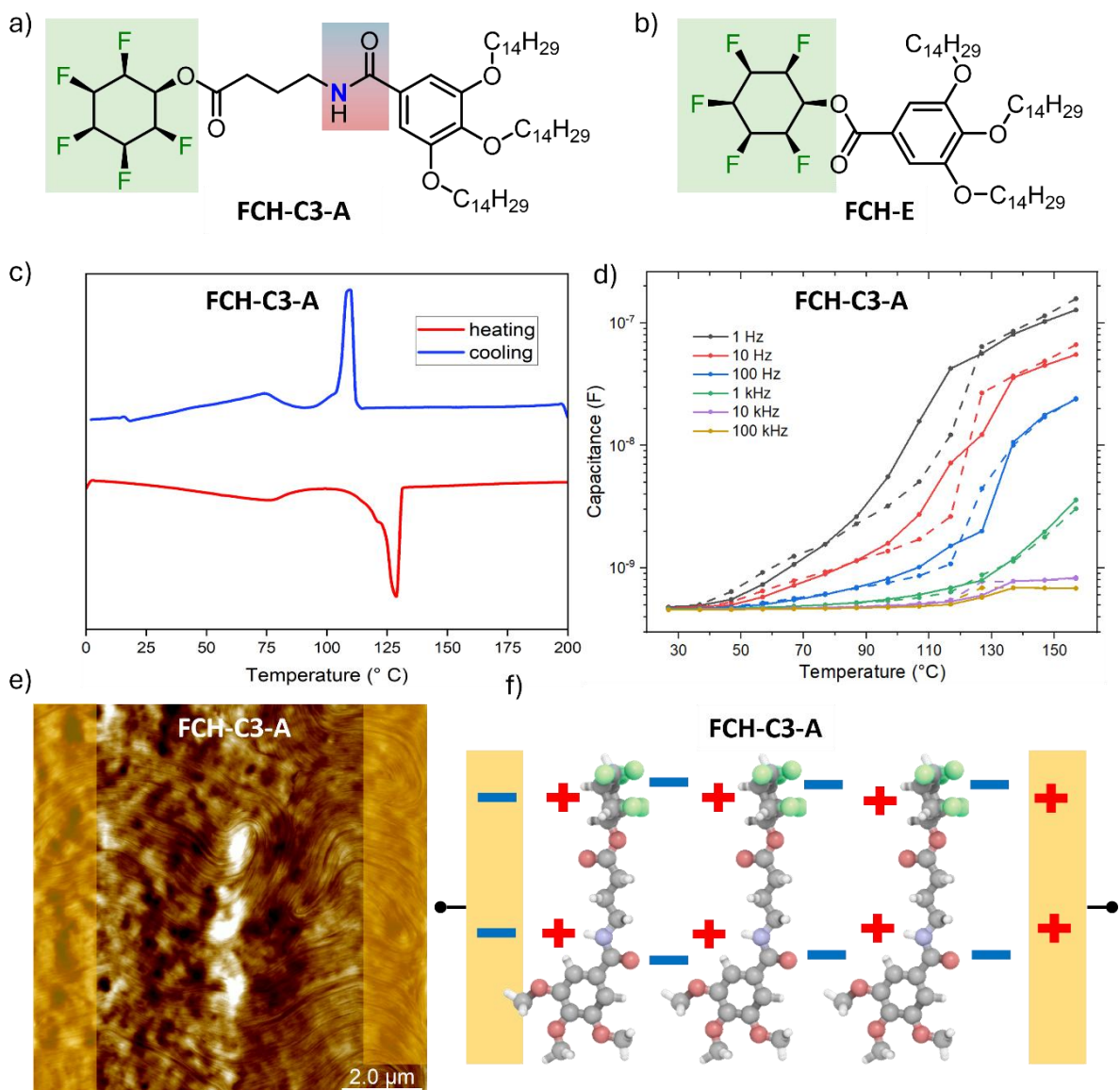

**Figure 1**: Molecular structures of (a) **FCH-C3-A** and (b) **FCH-E**. c) Differential scanning calorimetry traces of **FCH-C3-A.** d) Dielectric spectroscopy measurements at various frequencies, where solid lines signal heating and dotted lines cooling traces. e) AFM topography of annealed **FCH-C3-A** with the location of the buried metal electrodes indicated. f) Schematic illustration of supramolecular stacking of the building blocks within a **FCH-C3-A** fiber between electrodes, showing the plausible orientation of the two dipolar moieties. Red, blue and yellow indicate oxygen, nitrogen and fluorine atoms. Molecular structure optimized using the Universal Force Field (UFF).

As **FCH-C3-A** and **FCH-E** tend to form rough and not completely closed layers, and to facilitate topographical and structural investigations, patterned interdigitated electrodes (IDEs) with micrometer-sized gaps were used. Thin films were created by drop-casting and measured either in ambient conditions, in nitrogen atmosphere or in vacuum. In-plane electrodes and rough films make an exact quantitative measurement of current density and therefore polarization difficult. Hence only measured currents are shown and further dependent parameters are estimated. Likewise, the given electric field values (voltage/gap width) are approximate, as the in-plane geometry leads to bending of the field lines, which results in deviations in field strength compared to more symmetric out-of-plane parallel plate capacitor geometries. A detailed summary of sample fabrication, layout and characterization is given in SI section 2.

Ferroelectric materials commonly exhibit a phase transition from a higher temperature paraelectric phase to a lower temperature ferroelectric phase at the Curie-temperature, at which a structural transition from a state of higher symmetry to a state of lower symmetry occurs. The ferroelectric-to-paraelectric phase transition can be detected experimentally by various means, such as differential scanning calorimetry (DSC)[35], dielectric spectroscopy (DS)[33,35], second-harmonic generation (SHG)[36] and polarization switching measurements[37]. DSC measurements for **FCH-C3-A** are shown in Fig. 1c. Two endothermic peaks are observed in the heating trace, a small but broad one centered around 70°C and a larger one at around 127°C. The latter one is readily identified as the melting point of the material, while the smaller one implies another structural transition that will (below) be identified as associated with the mobilization of the amide group.

Ferroelectric materials commonly exhibit Curie-Weiß-like behavior of the electric permittivity, measurable by DS. Fig. 1d shows the measured capacitance, which is a sufficient proxy for the permittivity, of an **FCH-C3-A** coated IDE, depending on temperature for various frequencies. In contrast to the DSC measurement, we do not find peak-like features in the heating trace, instead a gradual increase of the capacitance with increasing temperature is observed, followed by a large jump to higher capacitances around the melting point. Since no Curie-Weiß singularity is observed at that temperature, we conclude that the DSC peak at 70°C is of non-ferroic nature, consistent with ferroelectric behavior found above 70°C, as discussed in detail below. The absence of Curie-Weiß-like behavior does not exclude ferroelectricity in materials, but implies that if proven ferroelectric, the Curie temperature of both compounds investigated here lies at temperatures at or above the melting point. This has been observed before, e.g. in P(VDF-TrFE) or in the high-temperature ferroelectric phase of $BaMnF_4$[38,39]. We explain the gradually increasing capacitance by a gradual mobilization of dipolar moieties with rising temperature, and therefore available thermal energy, in the system, which is in line with the fact that the capacitance upswing coincides with the DSC peak at 70°C. Another, not mutually exclusive possibility is the material approaching the transition to the paraelectric state but melting before reaching the transition temperature. At the melting point, a large enhancement in dipolar mobility leads to a steep increase in capacitance. Lower modulation frequencies (1-100 Hz) give rise to higher amplitudes already at lower temperatures, which implies relatively slow-moving dipoles that are difficult to excite at higher frequencies.

In the following, we present polarization switching measurements on films of **FCH-C3-A** and **FCH-E** using the double-wave-method (DWM)[37] that in principle allows extracting of the pure switching current by correcting for background currents caused by leakage and displacement. Subsequent integration then yields the ferroelectric polarization versus field loops.

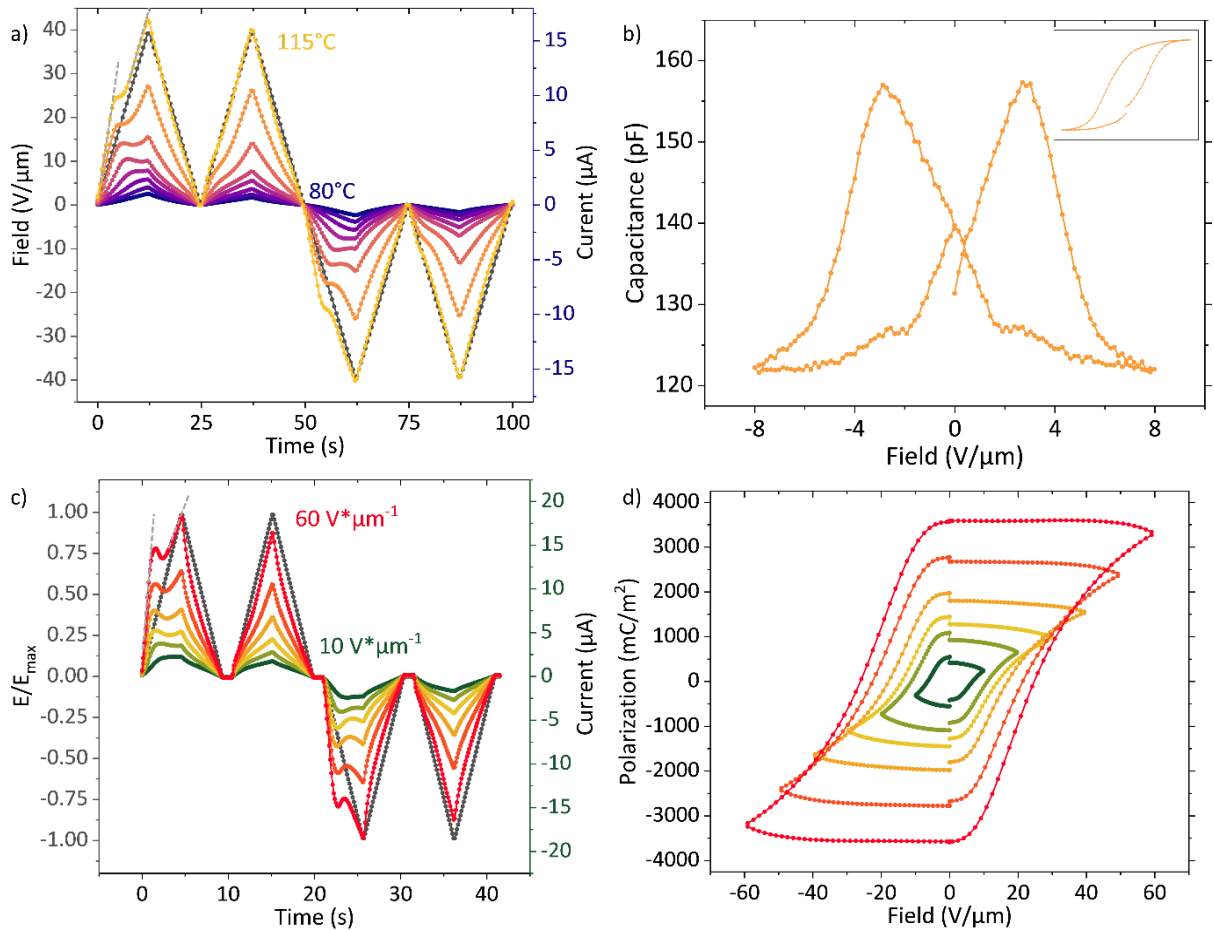

**Figure 2**: a) Double wave measurements at constant field (40 V/μm) and frequency (10 mHz) of **FCH-C3-A** for varying temperatures. The dashed grey lines are linear approximations of the slopes below and above the coercive field. b) shows a low field CV measurement at 110°C with the reversible part of the polarization obtained from integration shown in the inset; the vertical scale amounts to ±0.23 mC/m². c) Double wave measurements at constant temperature (100°C) and frequency (25 mHz) for varying applied fields in. d) "Polarization" hysteresis curves obtained by integrating the currents from c). The magnitude of the polarization and its non-saturating behavior indicates they are not true ferroelectric polarization loops as discussed in the text.

In Fig. 2a the response current to a DWM pulse at fixed amplitude and frequency is plotted for increasing temperatures. At around 90°C (shown in more detail in Fig. S1), a shoulder starts to appear in the first and the third readout pulse. With increasing temperature, the shoulder evolves into a pronounced peak, that progressively shifts its center towards lower fields. The most straightforward origin of such a peak is ferroelectric polarization switching. Alternative explanations in terms of accumulation of mobile ionic species at the electrodes or material degradation would not cause a peak but rather continue with increasing bias. Moreover, the presence of any significant number of mobile ions, as well as of degradation, has been excluded in previous work.[25]

The coercive field, which is the applied electric field where the bulk polarization completely inverts, is, in lowest order, identified as the field where the switching current is at its maximum. Ferroic switching in disordered organic molecules can typically be described as a thermally activated, nucleation limited switching (TA-NLS) process[40–42]. The dependence of the coercive field on temperature and sweep frequency for **FCH-C3-A,** shown in Fig. S2 and S3, shows qualitative agreement with the TA-NLS theory as well as with experimental data for the well-understood ferroelectric BTA.[34] The effect of the supramolecular fiber formation is seen in Fig. S4, where repeated DWM measurements lead to a continuously increasing conductivity, as well as a shift of the coercive field to higher fields. The latter can be explained in terms of the TA-NLS model by an increase in critical nucleation volume, associated with a reduced structural disorder.

DWM measurements for varying maximum applied external fields are shown in Fig. 2c. With increasing field strength, the switching peak takes form, decreases in width and its center is reached earlier in time. The peak center's shift comes to a stop at 30 V/μm, a further increase of field strength solely results in a slight narrowing of the peak and a strong increase in background current relatively to the peak amplitude. This peak evolution with field strength is typical for ferroic materials, as for fields below the coercive field only a subset of dipoles is flipped, according to the Preisach distribution of the material.[43] Once the coercive field is reached, all existing dipoles are flipped and the peak growth saturates, resulting in the saturation of the polarization when applying progressively increasing electric fields. However, the hysteresis loops associated with the currents from Fig. 2c show an ever growing "polarization" with field, as seen in Fig. 2d. Furthermore, the polarization values are multiple orders of magnitude above the theoretically estimated polarization of approximately 83 mC/m$^2$. This raises the question whether the material is truly ferroic, as leaky dielectrics can also show ferroelectric-like polarization loops.[44] However, the occurrence of current peaks at well-defined fields below the maximum applied field, and these peaks being describable in terms of the TA-NLS model, conflicts with the leaky-dielectric hypothesis.

A further sign of truly ferroic behavior can be found in capacitance-voltage (CV) measurements, where a DC-bias with a superimposed AC-modulation is swept back and forth, where the latter is used to measure the (small signal) capacitance. Ferroelectric materials possess a heightened dipolar sensitivity to external perturbations in vicinity of the coercive field, where the permittivity increases drastically. This results in a characteristic butterfly loop of the capacitance, with peaks in increasing field direction for both polarities. The coercive fields obtained from CV loops generally differ from the ones obtained from DWM measurement due to the different sweep rate for both experiments. A CV measurement of **FCH-C3-A** is shown in Fig. 2b and demonstrates the for ferroelectrics characteristic butterfly shape. Integrating the CV-loop yields a well-shaped hysteresis loop, reflecting the reversible polarization that is generally only a minor fraction of the total polarization.[11] Hence, we conclude that **FCH-C3-A** is a true ferroelectric. Therefore, the overestimation of the polarization has to result from the conducting properties of the **FCH-C3-A**, which apparently cannot be accounted for with the simple DWM correction that relies on the assumption that any background current does not depend on sample history and, as such, not on polarization. As discussed in a later section, the observed behavior is consistent with a polarization-modulated conductivity. Additional attempts of correcting for the conductivity as well as the corresponding hysteresis loops to the current measurements are presented in Fig. S5.

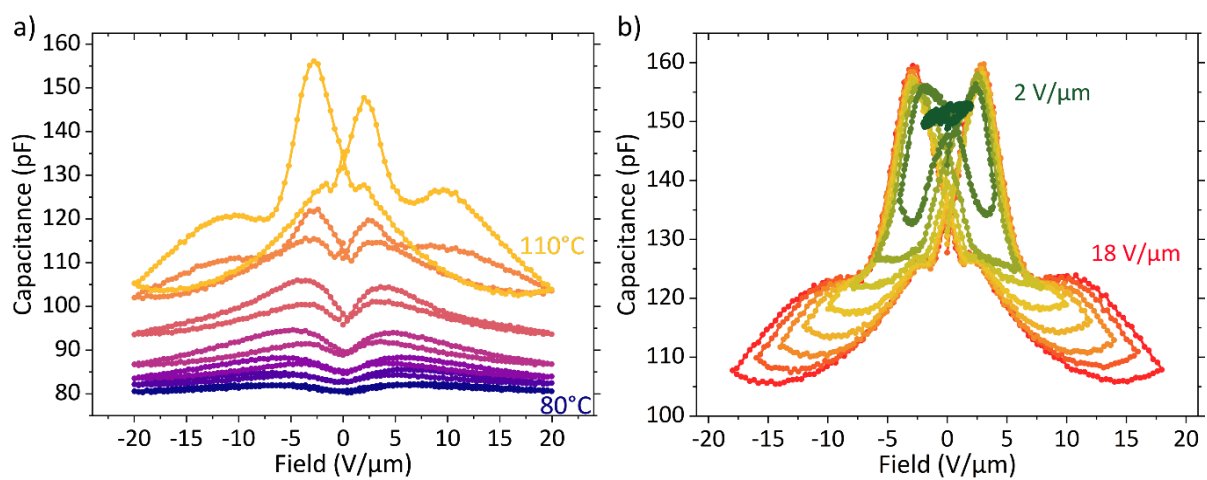

**Figure 3**: Capacitance-voltage measurements of **FCH-C3-A**, measured a) at fixed maximum DC-field (20 V/μm) for increasing temperatures ($\Delta T = 5°C$) and b) at fixed temperature (110°C) for increasing maximum DC-field strength ($\Delta V = 2$ V/μm). Both measurements are taken with a small signal frequency and amplitude of 100 Hz and 0.2 V/μm, respectively.

Additional insight into the switching process of **FCH-C3-A** could be obtained by sweeping a larger parameter space with CV measurements, see Fig. 3a. Around 85°C, the butterfly shape begins to emerge with its peaks becoming more pronounced and shifting to lower fields with rising temperature, in a similar fashion as in the DWM measurements. Interestingly, at 105°C a second feature appears on the flanks of either side of the butterfly loop, becoming more pronounced at 110°C. As a natural explanation for the occurrence of a second peak, the presence of two dipolar moieties, the all-*cis* fluorinated cyclohexane ring (FCH) and the amide group (A), in the **FCH-C3-A** comes to mind. This would require the spatially separated stacks of these groups to act as two, at least partially independent ferroelectric sublattices with separate coercive fields. To our best knowledge there is no existing literature regarding organic materials with two separate ferroelectric lattices, and there are no CV loops reported for such systems. Our CV loops are reminiscent of CV measurements on lead zirconium titanite (PZT) conducted by Bar-Chaim *et al.*[45] who attributed double peak butterfly loops to the presence of 90° and 180° domains, which possess different coercive fields.

The field dependence is investigated further in Fig. 3b, where CV curves for a stepwise increasing maximum applied DC-bias are depicted. The separate contributions of the two sublattices can be discerned. For small fields, incomplete loops are observed, as neither of the coercive fields is reached. From ~8 V/μm on, a well-formed complete butterfly loop is observed, and no further shift of the first peak is observed. The field of the peak maximum is the first coercive field with a magnitude of ~2.8 V/μm. At 10 V/μm the tails of the loops begin to broaden, eventually evolving into a complete second peak at ~14 V/μm and beyond. The corresponding coercive field to the maximum of the second peak is around 9.8 V/μm. Interestingly, once the second peak starts to develop, another small peak is observed around 1.9 V/μm in the direction of decreasing fields. A possible explanation is a field-dependent Curie temperature (of the second sublattice), in line with what has been observed previously for BTO and P(VDF-TrFE) and follows from the Landau-Devonshire theory of ferroelectrics.[46,47] For small fields and higher temperatures, these materials transition into a paraelectric phase, losing their polarization, which induces a peak in the CV curve for decreasing fields. However, as no corresponding features have been observed in the DWM measurements, we refrain from making a definite attribution.

Frequency dependent DWM measurements at 110°C displayed in Fig. S3 similarly show two separate peaks that can be simultaneously described by the TA-NLS model. The corresponding low-frequency coercive fields of 3 to 5 V/μm for the low-field peak and 10.5 to 12.5 V/μm for the high-field peak agree well with the coercive fields observed in the CV measurements. This finding reinforces the notion that both the peaks are of ferroic nature and can be attributed to the switching of the two dipolar moieties. An assignment of the individual peaks to a specific moiety is discussed further below. Additionally, in low-frequency, low-field DWM measured at 110°C shown in Fig. S6, it is possible to observe a transition from the first coercive field to the second, with a value of 1.7 V/μm for the former. Observing switching currents at two different fields qualitatively agrees with the findings from the CV loops, although it is difficult to discern at which fields exactly the transition from the switching of the first moiety to the switching of the second moiety occurs, which we attribute to a significant, disorder-induced overlap of the switching peaks, in combination with the background currents owed to the conducting property of the material.

The molecule **FCH-E** (see Fig. 2b) is structurally very similar to the **FCH-C3-A**, but is missing the amide group as well as the C3 spacer unit and, as such, might provide additional insight into the switching mechanism. In our previous work, **FCH-E** was shown to be an insulator, only showing small leakage currents, with field annealing not resulting in the formation of long-range supramolecular structures; 2D GIWAXS and high-resolution AFM measurements presented in Fig. S7 do show the presence of short-range stacking.[25] Moreover, as seen earlier in Fig. S4, a fully developed long-range supramolecular structure is no prerequisite for the ferroelectric properties of **FCH-C3-A**. In DSC and DS measurements on **FCH-E**, only melting and freezing points are observed (see Fig. S8). This suggests strongly that the peak at 70°C in the DSC measurement of **FCH-C3-A**, and concomitantly the low-field and low-temperature response of the latter material, are associated with the (mobilization of the) amide group.

Fig. 4a, b show temperature- and field-dependent DWM measurements of **FCH-E** for fixed sweep frequency. The resulting polarization loops are presented in Fig. S9a and d of the SI. Contrary to **FCH-C3-A**, the polarization shows indications for saturation at high fields and was limited to physically meaningful values, indicating a negligible contribution due to conductivity. The saturation polarization obtained from DWM measurements corrected for ionic contributions is around 5.1 ± 0.8 mC/m² (see SI Fig. S10), which is significantly lower than an approximate theoretical upper limit of 69 mC/m² for a fully flipped all-*cis* fluorinated cyclohexane group. This discrepancy suggests that only partial alignment of the dipolar moiety occurs, likely caused by insufficient structural ordering, causing off-axis dipole orientation or preventing switching altogether. Similar to **FCH-C3-A**, the temperature and frequency dependency of the coercive field are well described by the TA-NLS law, see Figs. S9c, d and S11c, d in the SI. A comparison of the fitting parameters with those obtained for the high field peak of **FCH-C3-A** show good agreement, reinforcing the assignment of the high field peak to the ferroelectric switching of the all-*cis* pentafluorocyclohexane group.

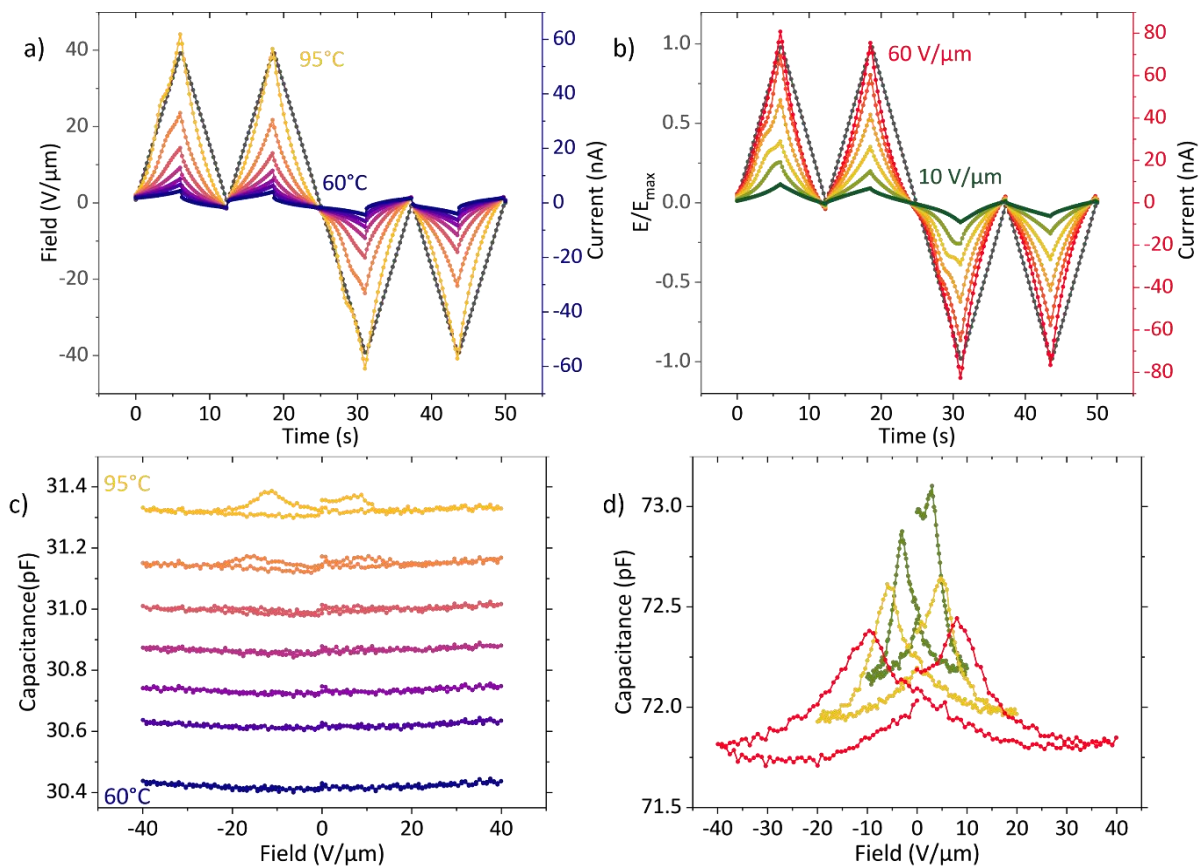

**Figure 4**: DWM measurements of **FCH-E** at fixed frequency and maximum field for increasing temperatures ($\Delta T = 5°C$) in a) and at fixed frequency and constant temperature of 95°C for increasing maximum fields ($\Delta E = 10$ V/μm) in b). The corresponding polarization hysteresis loops are shown in the SI Fig. S9. c) CV loops taken with a small signal peak to peak amplitude of 1 V/μm and small signal frequency of 1 kHz for increasing temperatures ($\Delta T = 5°C$). d) CV loops taken with a small signal peak to peak amplitude of 1 V/μm and a small signal frequency of 10 Hz for a stepwise increasing maximum field.

The temperature- and field-dependent CV loops of **FCH-E** are presented in Fig. 4c and d, respectively. In the former, ferroelectric butterfly loops are observed from 90°C on. The CV loops together with the DWM measurements are a strong indication of true ferroelectric switching in **FCH-E**. In the field dependent loops (Fig. 4d), the butterfly peaks continuously shift to higher fields with increasing applied maximum field. This is a result of the maximum applied field closing in on, and finally exceeding the coercive field, at which point the peak shift comes to a stop. Even at high DC-fields and low small-signal frequencies, only one peak is observed, supporting the hypothesis that the two peaks in the CV measurements of **FCH-C3-A** are caused by its two distinct dipolar moieties. Generally, the coercive fields for the **FCH-E** obtained from DWM and CV loops are around 20 and 10 V/μm, respectively. These values are of similar magnitude as the second, larger coercive field observed for the **FCH-C3-A**. This is consistent with the ascription made above that the smaller coercive field belongs to the amide group and, consequently, the larger coercive field belongs to the pentafluorocyclohexane group. This is also consistent with the intuitively expected lower steric hindrance for the re-orientation of the amide group.

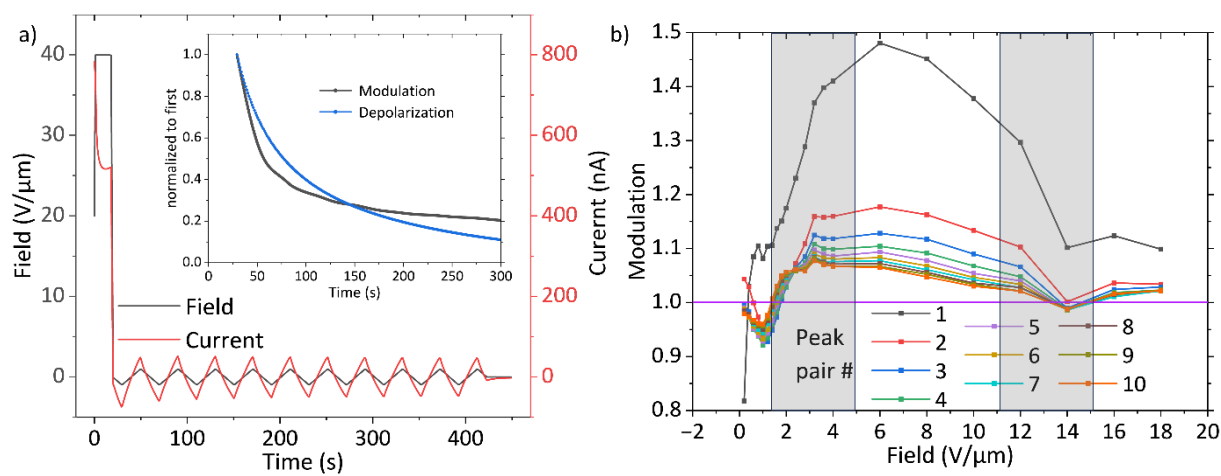

**Figure 5**: a) Schematic of the zig-zag conductivity switching measurements on **FCH-C3-A** films, where one rectangular poling pulse is followed by triangular pulses of alternating sign. These measurements correspond to one data point per peak pair in image b), where the absolute peak ratio (conductivity modulation) for each peak pair is plotted against the triangular voltage amplitude. The overlaid gray bars indicate the approximate ranges of the coercive fields obtained from DWM and CV measurements. The inset to a) plots the normalized current modulation together with the normalized depolarization current from the 40 V/μm poling pulse versus time.

In the final part of this article, evidence for polarization-mediated switching of the charge injection and bulk conductivity in **FCH-C3-A** is discussed. To this end, we employed a triangular bipolar "zig-zag" pulse sequence, where a single rectangular poling pulse is followed by multiple triangular pulses of alternating sign, therefore probing along and against the polarization direction. The conductivity modulation at the $n^{\text{th}}$ pair of negative and positive voltage peaks can then be calculated as the ratio of the corresponding absolute currents. A detailed showcase of the process including background current correction is presented in the SI in Fig. S12 and an alternative approach to eliminate the time delay between peaks via extrapolation in Fig. S13. Any conductivity modulation is expected to vanish for triangular amplitudes above the coercive field. Likewise, linear response theory demands that no modulation occurs in the low-field regime as the derivative of the current-voltage characteristic must be well-defined at zero voltage. In addition, the presence of two different coercive fields has to be kept in mind. Fig. 5a shows a schematic of the zig-zag measurement, which was done for varying maximum triangular amplitudes.

To start with, the effects of depolarization, that is the relaxation of the dipolar ordering over time, have to be considered. Depolarization of the material is greatly accelerated at 100°C and results in a current of opposite sign compared to the poling pulse, which is included in the measured currents and hence is superimposed on the possible conductivity modulation. The associated currents were previously shown to be small in the used materials; the same holds for any spurious ionic contributions to the current transient.[25] The current transient due to depolarization and any ions can be extracted from the zero crossings of the zig-zag measurements and is shown in Fig. S12 for the 1.8 V/µm triangular maximum field. After correcting for these background currents, the ratio of each negative triangular pulse peak to its successive positive counterpart is taken and plotted in Fig. 5b against the applied maximum triangular field, for all peak pairs. Over the change from the first to the last peak pair, the effect of the polarization decay is observed, resulting in the decrease of the polarization modulated conductivity. Since the depolarization current falls to 50% of its starting value in ten seconds, see Fig. S12c, the modulation concomitantly strongly decreases over time. In the inset of Fig. 5a, the normalized depolarization current and the normalized modulation are plotted in parallel, showing a clear correlation between (de)polarization and conductivity modulation, suggesting a causal relation.[48]

As mentioned above, the bulk conductivity modulation is a higher order effect and should vanish at low fields, i.e. converge to a value of unity. Surprisingly, prior to turning positive, the modulation first becomes smaller than one. We attribute this to injection barrier modulation (IBM), where currents in the polarization direction are increased and which experimentally was observed at low fields.[21,26] Irrespective of that, the successive increase in modulation to values over unity is indicative of bulk conductivity switching (BCS), where the conductivity is increased when field and polarization are antiparallel and vice versa.[21,48] As such, IBM and BCS are competing effects and it was shown that in materials exhibiting both effects IBM is observed at low fields, while BCS begins to dominate at intermediate to high fields. In our previous work, we demonstrated the existence of a substantial injection barrier between the Au electrodes and the active material, leading to the necessity to (locally) align the molecular dipoles with the electric field at the injecting contact in order to obtain an appreciable, and potentially non-limiting charge injection.[25] Hence, some form of IBM is expected at low field. Note that the coercive field is a bulk material property and that the reversal of individual dipoles in the likely more disordered contact regions is expected to occur already at substantially lower fields[49].

To substantiate the assignment of the low-field negative conductivity modulation to IBM, we performed in-operando Kelvin probe force microscopy (KPFM) measurements of the (spatial distribution of the) electrostatic potential in the channel. The results are shown in Fig. 6. For finite applied fields, we observe step-like features at both contacts that are superimposed on a more or less linear background. At low fields, the voltage steps increase linearly with field and are of the same magnitude at both contacts, as expected for field-induced interfacial dipoles. For the shown fields, their magnitude remains below the estimated injection barrier between Au and **FCH-C3-A** of ~1.7 eV[25]; instrument limitations do not allow to measure beyond 2 V/μm (+/-10V), preventing the direct observation of the expected saturation of the interfacial dipole, either at the value needed to make the contact Ohmic, or at the value set by the available dipoles. The former and latter would be slightly below 1.7 eV and around 2 eV, respectively.[25] Importantly, the measured interfacial energy shifts are of a magnitude, and occur in a voltage range, that is fully consistent with our interpretation of the negative current modulation in Fig. 5b in terms of injection barrier modulation.

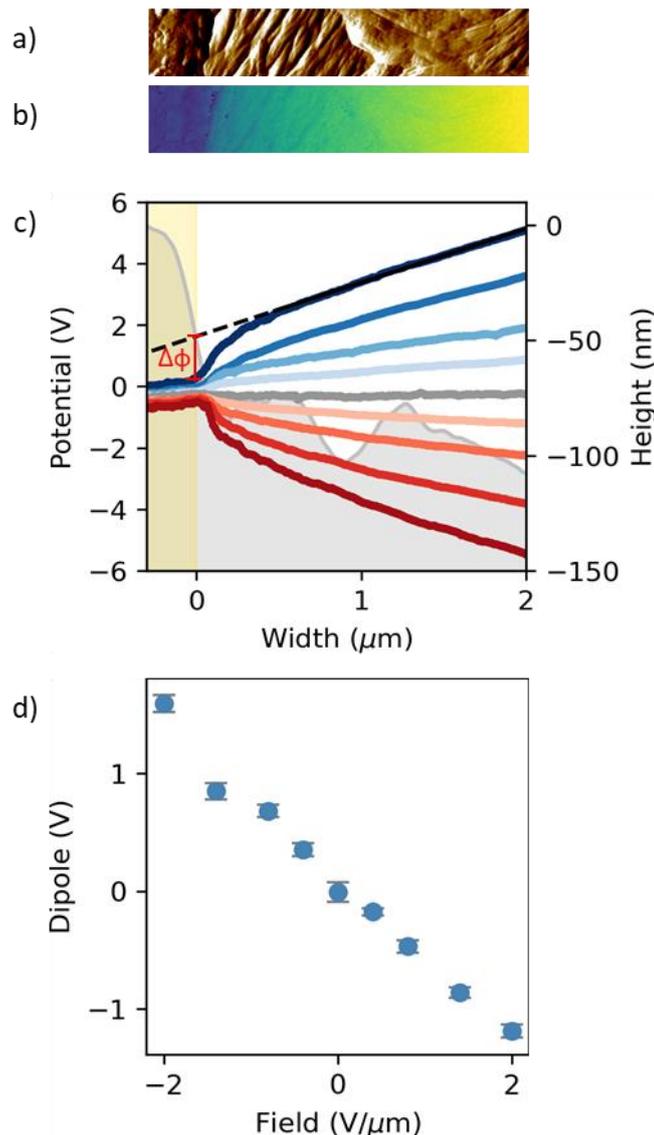

**Figure 6**: In-operando KPFM measurements on a pristine **FCH-C3-A** device with 5 μm inter-electrode gap. a) Amplitude signal, showing the edge of the (biased) electrode on the left-hand side; the (grounded counter electrode is not shown. b) Corresponding surface potential measured in sideband mode. c) Line sections of surface potentials at different applied voltages between +10 V (red) and -10 V (blue). The yellow shade indicates the electrode position. d) Interfacial dipole vs. field $\Delta\phi$, extracted as illustrated in c).

In Fig. 5b, beyond ~2 V/μm, depending on peak pair, the steady decrease in modulation with increasing field can be explained by bulk conductivity modulation with a continuously increasing fraction of reversed polarization. A further, significant drop to unity is observed at 14 V/μm, which is in the range of the second coercive field. From here on, the polarization is switched back and forth every triangular pulse and the modulation vanishes. The peak ratio being slightly above unity beyond the coercive field can be explained by polarization imprint effects, where a portion of hysterons becomes permanently trapped in one state, which can result from the asymmetric poling procedure.[50] Importantly, the qualitative shape of Fig. 5d, and particularly the (near) vanishing of the modulation beyond the highest coercive field excludes more trivial explanations in terms of, e.g., ionic currents.

Although **FCH-C3-A** is not a conventional organic semiconductor in that it lacks an extended π-system and charge transport accordingly involves a more general mechanism of subsequent oxidation and reduction of the energetically most accessible state, the observed behavior fits well to observations and simulations of BCS in 'conventional' semiconducting organic ferroelectrics by Gorbunov *et al.* and Casellas *et al.*[21,27] The former modeled BCS as Marcus hopping between discrete states in an asymmetric potential landscape, modulated by the ferroelectric polarization. The mechanism was later expanded by Johann *et al.*, combining molecular dynamic simulations and density functional theory[21,29]

An alternative and potentially more direct way of estimating the conductivity modulation is by taking the ratio of the slopes of the current response vs increasing field in the first (anti-parallel current and polarization) and second (parallel current and polarization) peaks of the DWM measurement, cf. the dashed grey lines in Fig. 2a, c. This modulation factor, by its definition, is also the modulation factor required to explain the abnormally high "polarization" in Fig. 2d. The modulation data obtained from the current slopes as well as the corresponding fits are shown in Fig. S14. Disregarding low fields and temperatures, where incomplete switching results in reduced slopes after the coercive field and therefore exaggerated modulation factors, the saturated on/off-ratios are between 2 and 5, depending on the exact measurement parameters. These exceed the maximum modulations in Fig. 5b and Fig. S13d that are around 1.3 to 1.5. The discrepancy can be explained by the effects of depolarization affecting the zig-zag measurement. The depolarization current, see Fig. S12c, reduces by a factor ~3 after 30 seconds, implying, in lowest order (cf. inset of Fig. 5a), a reduction of the remaining polarization by a similar amount. Hence, the depolarization would reduce the on/off-ratios obtained from the DWM slopes to ~1.3-2.3, comparable to those of the zig-zag measurements.

**Summary**

In this work, we investigated the electronic properties of the multi-functional small-molecule material **FCH-C3-A**, which, in solid state, forms quasi-1D supramolecular bundles. Combining

CV and DWM measurements, we showed that thin films of this material have ferroelectric properties, brought about by two largely independent polar lattices that are formed by spatially separated sub-stacks of the two dipolar groups in the **FCH-C3-A** material. The two sublattices are characterized by different coercive fields and accordingly show different temperature and frequency dependencies. Similar measurements on a molecular derivative **FCH-E** with only a single dipolar moiety further confirmed the ferroelectric properties of this family of materials and allowed an assignment of the coercive fields to the specific dipolar moieties. Integrating up the apparent switching currents obtained in the DWM measurements led to abnormally high and non-saturating "polarization" hysteresis loops, with "remnant polarization" values far exceeding the dipolar density of the material. This behavior was explained in terms of a modulation of the conductivity by the ferroelectric polarization. As it was previously shown that charge transport in **FCH-C3-A**, which lacks an extended π-system, differs from that in conventional organic semiconductors, this finding generalizes the polarization-modulation of the electrical conductivity, which was previously reported in conventional organic semiconducting ferroelectrics, to this new class of conducting materials.

**Acknowledgements**

We thank the Deutsche Forschungsgemeinschaft (DFG, German Research Foundation) for support of this work (project numbers 281029004 (SFB 1249) and DE-1830/6-1). M. K. thanks the Carl Zeiss Foundation for financial support.

Supplementary Information to
**A supramolecular ferroelectric with two sublattices and polarization dependent conductivity**

H. Mager[1], M. Litterst[1], Sophia Klubertz[1], Shyamkumar V. Haridas[2], Oleksandr Shyshov[2], M. von Delius[*,2] and M. Kemerink[*,1]

[1] Institute for Molecular Systems Engineering and Advanced Materials, Heidelberg University, Im Neuenheimer Feld 225, 69120 Heidelberg, Germany

[2] Institute of Organic Chemistry, University of Ulm, Albert-Einstein-Allee 11, 89081 Ulm, Germany

*Corresponding author e-mail: max.vondelius@uni-ulm.de; martijn.kemerink@uni-heidelberg.de

## Contents

## 1 – Materials

Full compound names and structures:

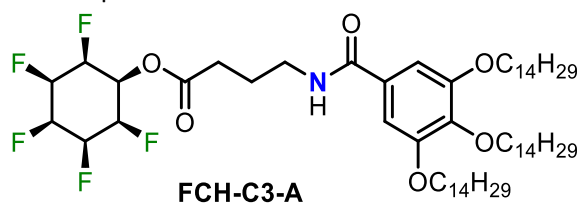

**FCH-C3-A**

**FCH-C3-A** is (1r,2R,3R,4s,5S,6S)-2,3,4,5,6-pentafluorocyclohexyl 3-(3,4,5-tris(tetradecyloxy)benzamido)butanoate

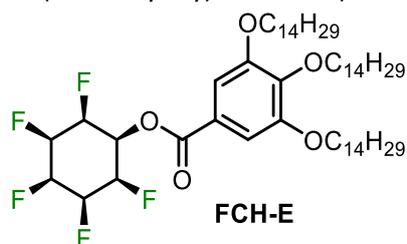

**FCH-E**

**FCH-E** is (1r,2R,3R,4s,5S,6S)-2,3,4,5,6-pentafluorocyclohexyl 3,4,5-tris(tetradecyloxy)benzoate
Materials were synthesized as described elsewhere.[1]

## 2 – Sample fabrication and methods

*Substrate preparation and characterization:* The thin film characterization was done on interdigitated electrodes (IDEs) which were either supplied by MicruX Technologies IDE or made by ourselves. For the former, the 180 pairs electrode pairs consist of 150 nm gold on top of 50 nm titanium deposited on a glass substrate. The total channel width amounts to 0.8 m. The latter IDEs are produced by a UV photolithography process inside a cleanroom and consist of typically 25-30 nm gold on top of 3 nm chromium, both thermally evaporated on a glass substrate. The total channel width amounts to 4 m. In both cases, the electrodes were 5 µm wide with a gap of 5 um between the electrodes.

Prior to active layer deposition, all substrates were cleaned the same way. The substrates were first mechanically washed with soap and water. Then, they were spaced out and put in succession into water, acetone and isopropanol for chemical cleaning. For each solution, the substrates were cleaned for 10 min at room temperature using an ultrasonic bath. Finally, the substrates were blown dry with nitrogen.

*Film deposition:* For reasons of material conservation, all samples were drop-casted from solution. **FCH-C3-A** and **FCH-E** were heated over the melting point and cooled back down before any measurement was taken. This was done to improve film homogeneity. **FCH-C3-A** and **FCH-E** were dissolved in tetrahydrofuran (THF) with concentrations between 10 to 20 mg/mL and drop-casted on glass substrates patterned with IDEs. To ensure full electrode coverage, 2-5 times 4-5 µL were drop-casted as needed. No additional temperature treatment was used during the deposition process.

All samples had full electrode coverage with inhomogeneous films with film thicknesses typically ranging from 1 to 3 µm.

The capacitance-voltage (CV) and double wave method (DWM) measurements were taken using an aixACCT Systems Research Line DBLI together with a Linkam HFS600E-PB4 probe stage. For the CV measurements, a DC-bias is swept back and forth with a superimposed small AC-bias to measure the capacitance.

*Surface characterization:* Atomic force microscopy (AFM) images were obtained using a Bruker MultiMode 8-HR AFM operated in tapping mode.

*Kelvin probe force microscopy*: KPFM images were taken with an Oxford Instruments Jupiter XR Asylum Research AFM in sideband mode together with a Linkam HFS600E-PB4 probe stage for temperature control and device contacting. Voltages were applied by a Keithley 2636B System Source Meter.

*Structural analysis:* The XRD measurements were carried out using a Rigaku SmartLab setup (Cu *source* with a 0.154 nm wavelength). The GIWAXS measurements were done in 2D mode with an incident angle of 0.30° and an exposure time of 3600 s.

*Differential Scanning Calorimetry*: DSC was conducted on Mettler Toledo DSC 2 STARe system under nitrogen atmosphere (heating and cooling was carried out at a rate of 5 °C/min).

## 3 – Additional measurements and analysis

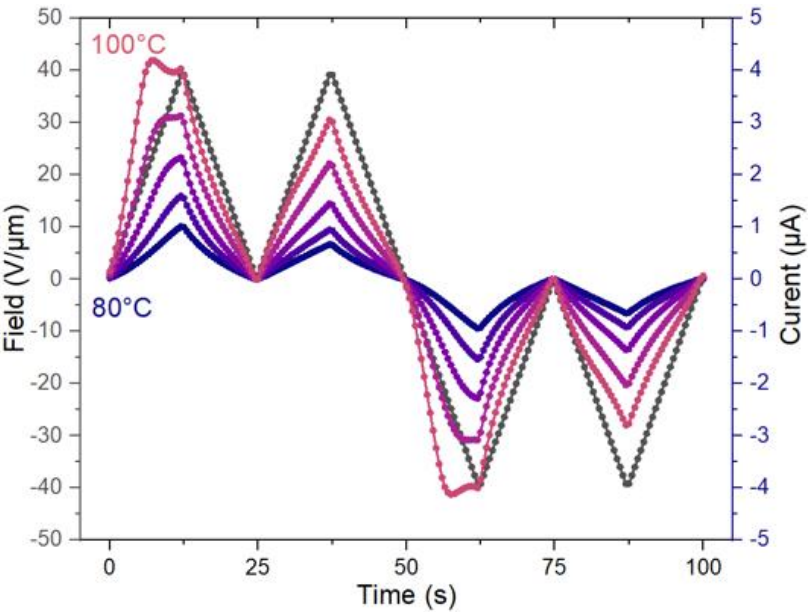

**Figure S1**: Zoom-in on lower temperatures in Fig. 2a, showing the emergence of the ferroelectric switching peak of **FCH-C3-A** at 90°C.

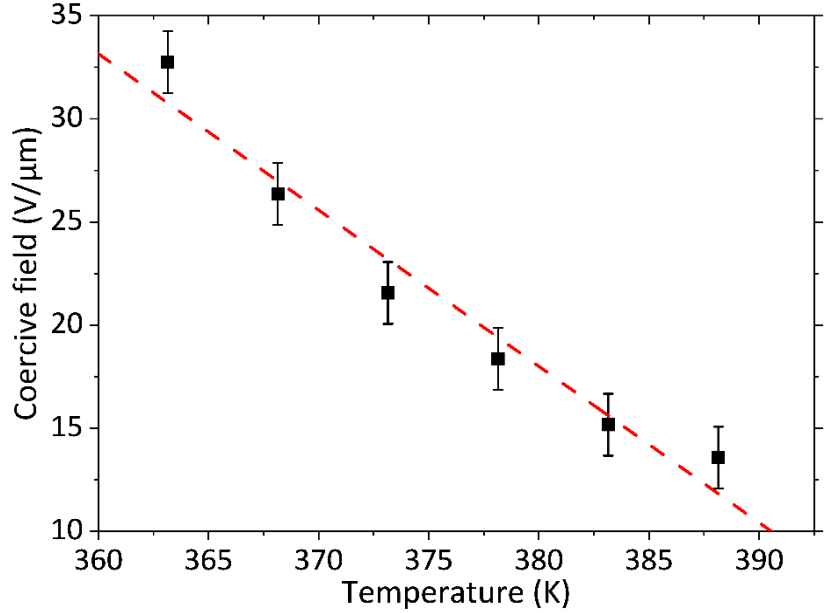

**Figure S2**: Coercive field values of **FCH-C3-A** (at 40 V/µm and 10 mHz) obtained from the field at the peak current in Fig. 2a plotted against temperature. The data is fitted with the TA-NLS model as described below.

Building on the Kalmogorov-Avrami-Ishibashi (KAI) model[2], Vopsaroiu *et al.* developed a model for polarization switching in ferroelectrics, based on thermally activated, nucleation limited switching (TA-NLS)[3,4]. For the coercive field, it gives the following dependence:

$$E_c = \frac{w_b}{P_r} - \frac{k_b T \cdot \ln(\nu_0 \tau \cdot \ln(2)^{-1})}{P_r V^*} \qquad 1$$

Here $w_b$ is the activation energy density of the critical nucleus required for polarization switching and $V^*$ its volume. $\nu_0$ is an attempt frequency, typical on the order of phonon frequencies of the material, $P_r$ the remnant polarization and $\tau$ the rise time of the switching pulse. $w_b$ and $V^*$ are taken as fitting parameters, with an attempt frequency of 10 THz that is on the order of typical vibration frequencies in such materials. As explained later in the main text, the high field peak in the DWM measurements is assigned to the pentafluorocyclohexane group, resulting in a theoretical $P_r$ value of 52 mC/m² and $\tau$ corresponds to 12 s for the measurement. This results in a $w_b$ of $99 \pm 1$ meV/nm³ and a $V^*$ of $11 \pm 1$ nm³. Both values are reasonable and are of similar magnitude as for the liquid crystalline BTA or the polymer P(VDF-TrFE).[5,6] It has to be noted, though, that $w_b$ and $V^*$ are in reality often better represented by distributions instead of by fixed parameters.[6] In addition, domain growth as described by the KAI model can be combined with the TA-NLS model for a more complete description. While incorporating the parameter distributions and the KAI model might correct for minor deviations from the TA-NLS model seen in our data, it is beyond the scope of this article.

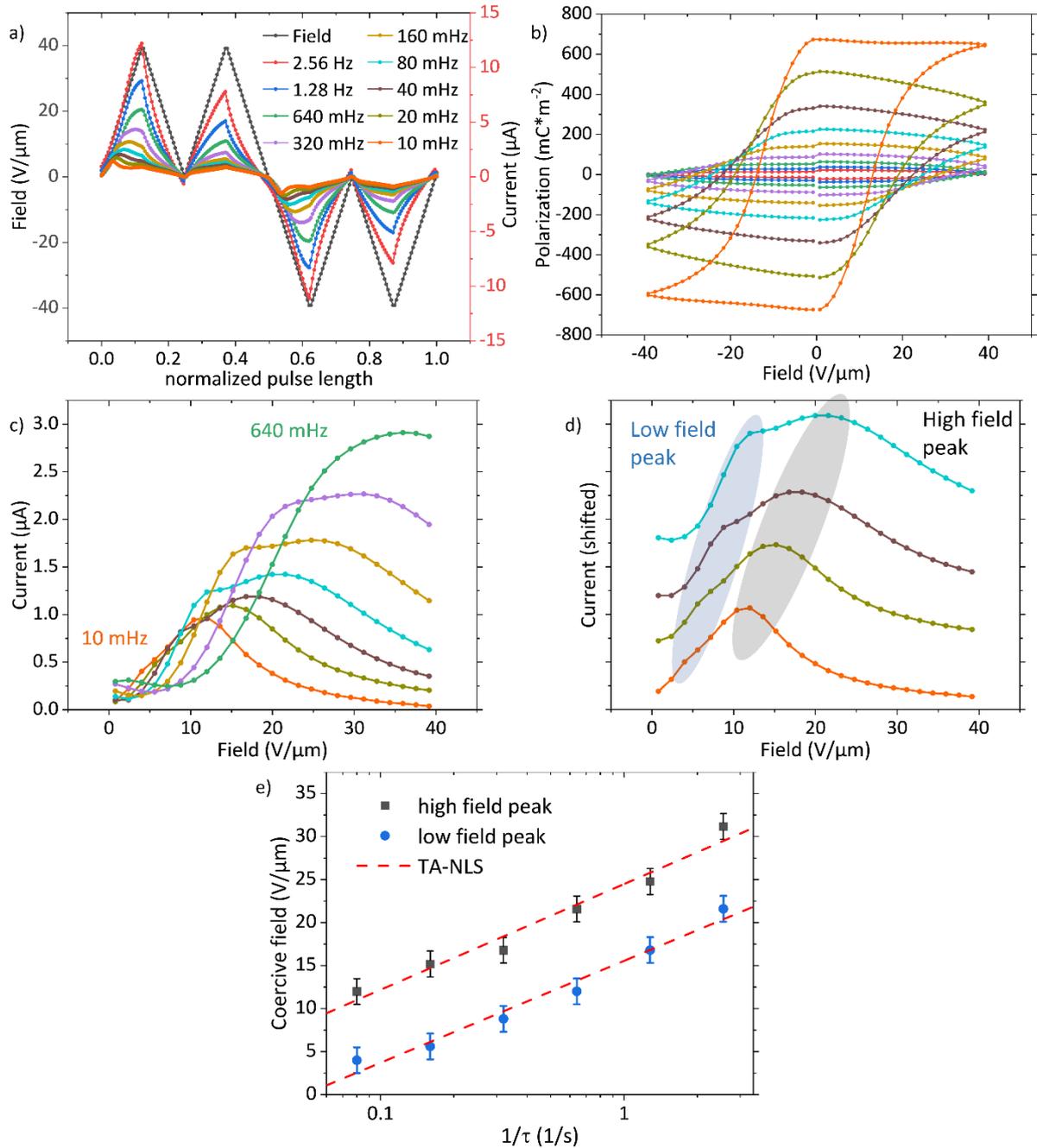

**Figure S3**: a) Frequency-dependent DWM measurements of **FCH-C3-A**, measured at 110°C. The x-axis is normalized to the pulse length. The lowest measurement frequency is 10 mHz which is then doubled for every measurement. Above 80 mHz parts of the switching peak are cut off. b) shows the corresponding hysteresis loops. c) shows the current transients obtained after DWM correction of a) and d) is a zoom-in to lower fields where two separate peaks are clearly visible. e) depicts the two coercive fields obtained from c), plotted against the inverse rise time, fitted with the TA-NLS model (equation 1). The fixed fitting parameters are an attempt frequency of 10 THz, a temperature of 383.15 K and the theoretical polarization values of 31 mC/m$^2$ and 52 mC/m$^2$, for the low and high field peak, respectively. Comparisons to the coercive field behavior of FCH-E and with the TA-NLS fitting parameters obtained in Fig. S11 indicate that the low field peak belongs to the amide group and the high field peak to the pentafluorocyclohexane group. The obtained fitting parameters for the low field peak are an energy barrier $w_b$ of $33.1 \pm 0.3$ meV/nm$^3$ and a nucleation volume $V^*$ of $33 \pm 3$ nm$^3$. For the high field peak $w_b$ of $60.4 \pm 0.5$ meV/nm$^3$ and a nucleation volume $V^*$ of $19 \pm 2$ nm$^3$ are obtained. The high-field peak values are reasonably consistent with the values found for the temperature dependence in Fig. S2.

Since the peaks in the DWM are conductivity peaks and not switching peaks, the lack of a visible low field peak at lower temperatures in the DWM (cf. Fig. 2a) is explained by the presence of a finite injection barrier, which is only overcome once the pentafluorocyclohexane group becomes mobile as well. This is consistent with the injection barrier limited behavior seen at low fields in Fig. 5b. If this case, the amide switching peak should be observed instead. The reason the amide switching peak is not detected can be explained by the fact that compared to the general magnitude of the measured current, the amide switching current portion is small and therefore drowned out (cf. Fig. S1).

In contrast, the CV measurements depicted in Fig. 3 show actual reversible ferroelectric switching, therefore the low field peak of the amide group is already visible at lower temperatures, independent of injection barriers. Reversible switching of the pentafluorocyclohexane group is assumed to be energetically costlier, meaning higher temperatures/fields and lower frequencies are required.

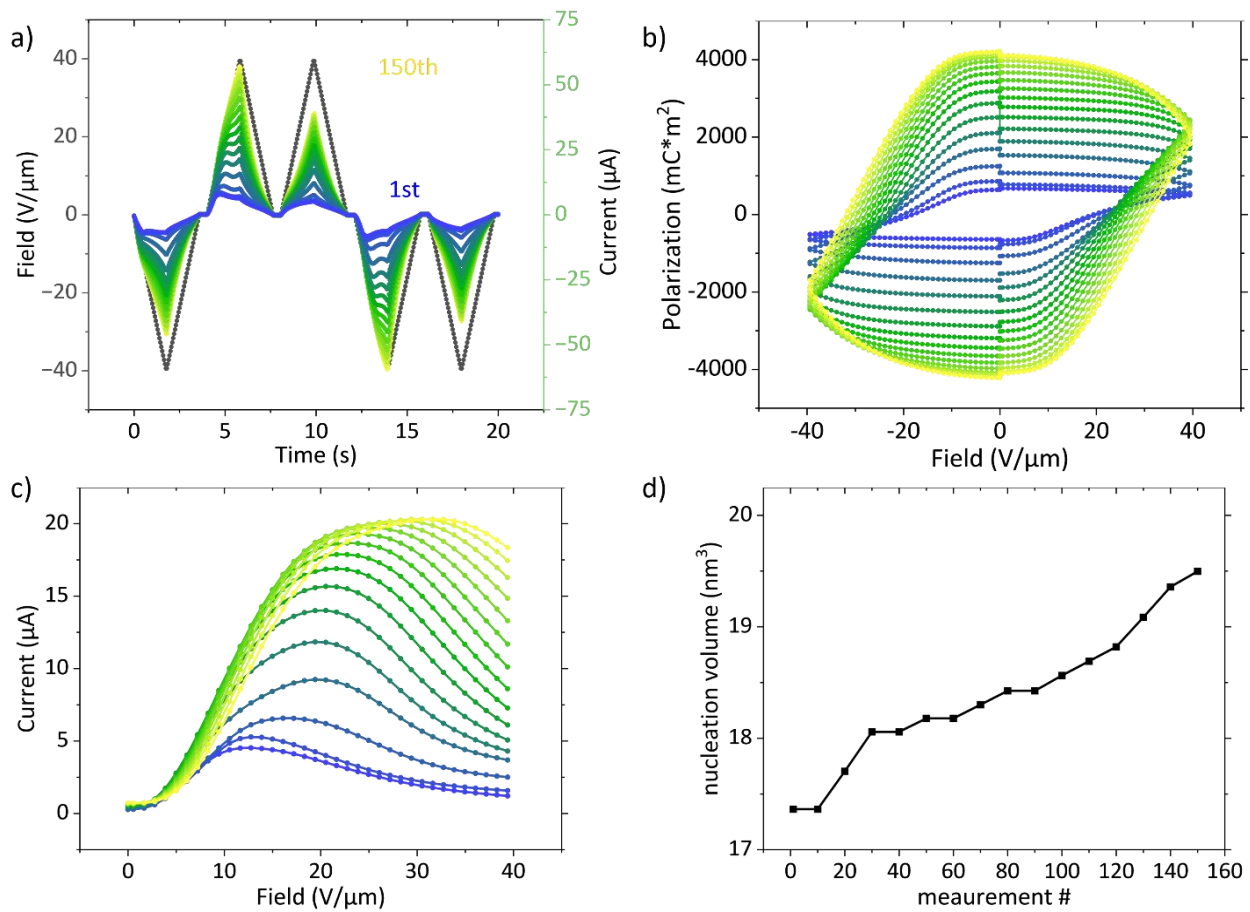

**Figure S4**: a) 150 consecutive DWM measurements of **FCH-C3-A** at fixed parameters of 60 mHz, 100°C and 40 V/μm. Only every tenth measurement is plotted. b) Corresponding "polarization" hysteresis loops, which keep growing as a result of the increasing conductivity. d) Obtained current transients after DWM background correction. c) Critical nucleation volume calculated for the coercive fields obtained and an energy barrier $w_b$ of 60 meV/nm$^3$ obtained via the TA-NLS model.

Fig. S4 shows the effect of 150 consecutive measurements at fixed field, temperature and frequency on the switching and conducting properties. Two significant changes in the current response can be observed. On the one hand, the switching peak shifts to higher electric fields/later timestamps while on the other hand the background current increases by approximately one order of magnitude. As a result, part of the (apparent) switching current is cut off in later measurements, a higher field or lower frequency is required for full polarization switching at this point. The concomitant strong increase of the background current starts drowning out what remains of the switching peak. Irrespective of the (polarization-dependent) conductivity mechanism, the explanation for these trends can be traced to the morphological change of the material during measurements. In previous work, the **FCH-C3-A** was shown to form supramolecular fibers during field annealing (see Fig. 1e), which was accompanied by a significant rise in conductivity, followed by saturation and a soft roll-off.[1] While the fiber formation enhances material conductivity and therefore the background currents in DWM measurements, the shift of the switching peak towards higher fields additionally suggests that at the same time the dipole switching is progressively made more difficult. This can be readily explained in terms of the TA-NLS model, where with increasing order and supramolecular structure, the critical nucleation volume increases, which in turn requires progressively larger electric fields to overcome the potential barrier between ferroelectric states. The evolution of the critical nucleation volume obtained from the TA-NLS is shown in Fig. S4d. In short, as its structural order increases, the material transitions to slightly more intrinsic (but still extrinsic) switching behavior. Melting the material reverts the structural changes and resets the conducting and ferroic properties. While the change in ferroelectric and conducting properties allows for a certain tunability, it also makes an unambiguous determination of the ferroelectric characteristic parameters difficult, as they are heavily dependent on the sample's current morphology and therefore measurement history. Fully aligning to obtain a comparable base is not an option, as the resulting conductivity overshadows the ferroelectric effects. For comparability's sake, all data shown is collected from more or less pristine samples.

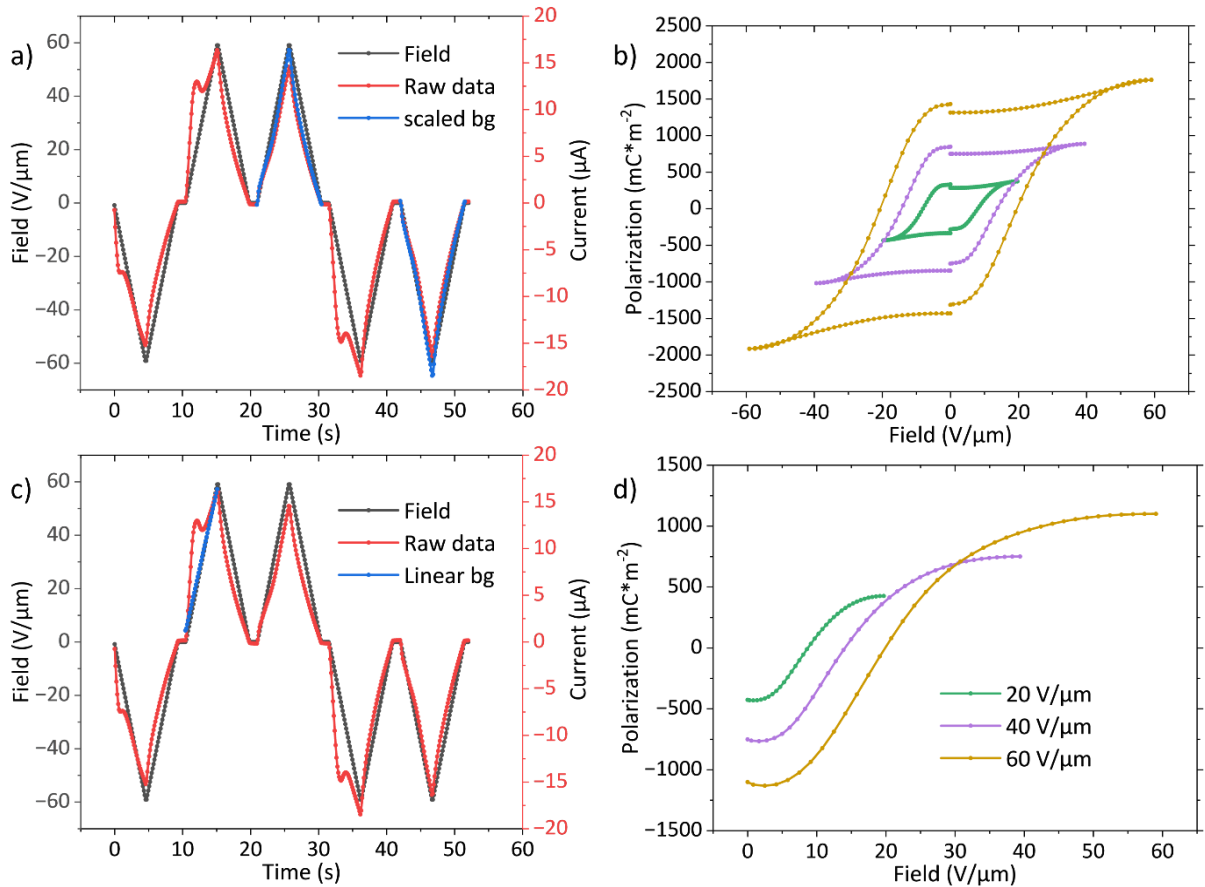

**Figure S5**: Different ways of background current correction are shown in an attempt to extract the contribution by true polarization reversal from the total apparent switching current. In a) a linear background current is subtracted from the current of the rising flank with the resulting integrated charge plotted on the right. As the resulting charge still vastly exceeds to theoretically expected polarization charge, this background correction is insufficient. Going one step further, the non-switching peaks in b) are scaled to the same value as the background current peak in the switching peaks. This should account for transient current effects. Subtraction of the scaled peaks from the switching peaks and integrating gives the "polarization" plotted on the right. Although reduced, its magnitude is still too large to be the result of dipolar switching.

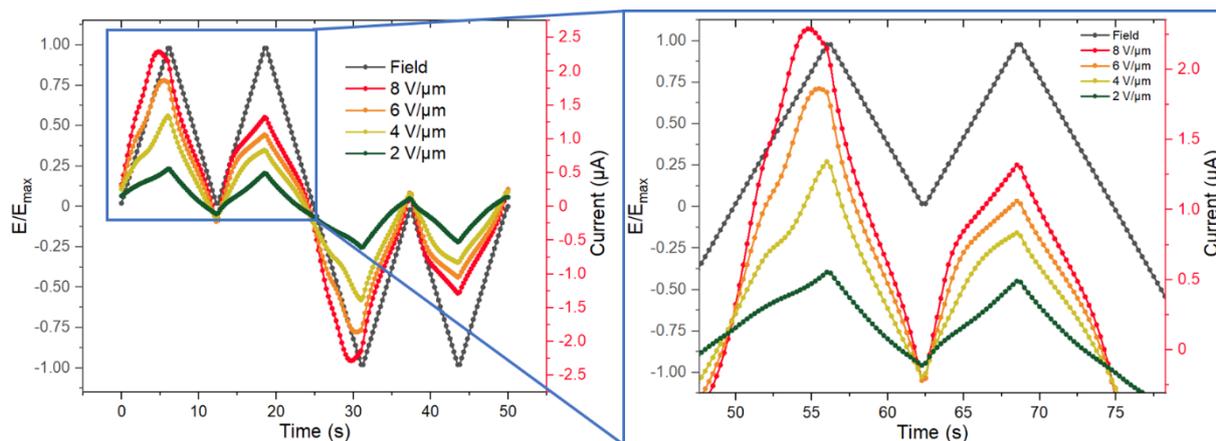

**Figure S6:** Low field DWM on **FCH-C3-A** at 110°C and 20 mHz. In the zoom-in on the right two shoulders can be clearly seen in the first pulse of the 6 V/μm line. We attribute the presence of two shoulders to the partially separate switching of the two dipolar moieties of the molecule, that is also observed in the CV measurements in Fig. 3 and in the frequency dependent DWM in Fig. S3. The shoulder in the second peak is attributed to incomplete polarization switching in the first peak.

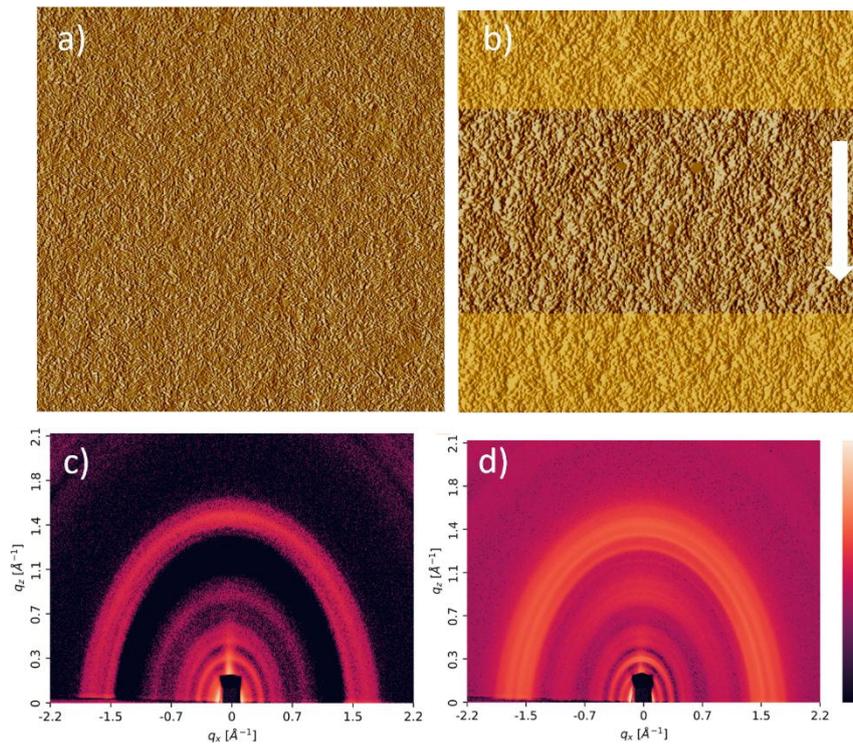

**Figure S7**: a) and b) show atomic force microscopy amplitude channel images of a **FCH-E** thin film before and after annealing at 90°C for one hour with an applied field of 20 V/μm. The buried electrode structure is indicated by the golden bars and the white arrow shows the field direction. c) and d) show corresponding grazing incident wide angle scattering images.

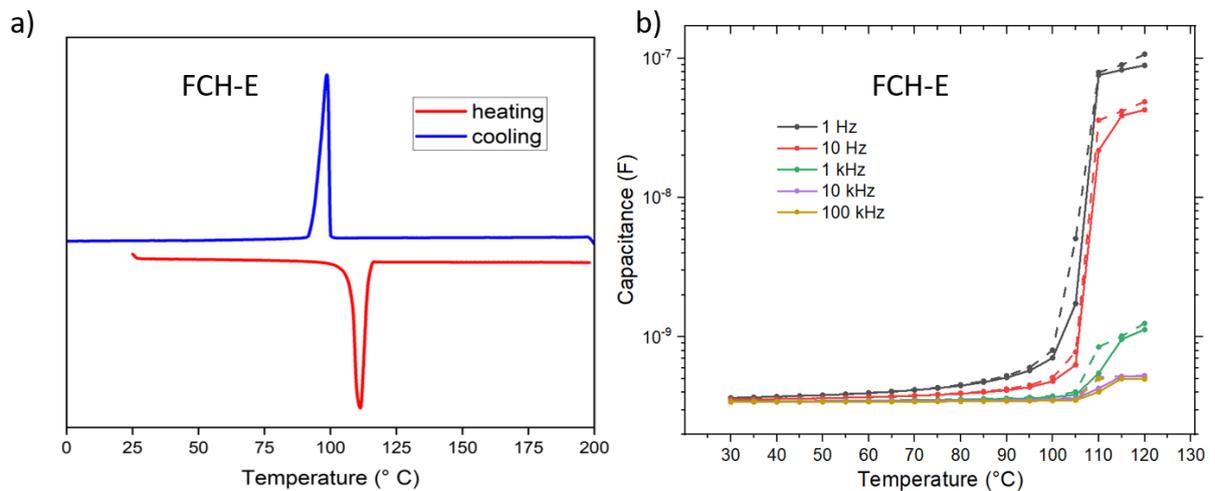

**Figure S8**: Phase behavior of **FCH-E**. a) DSC traces. Only one peak at around 110°C in the heating trace associated with the melting of the material is observed. The corresponding peak of the freezing point is at 95°C. b) Dielectric spectroscopy measurements, exhibiting a stepwise capacitance increase at the melting point and no other features.

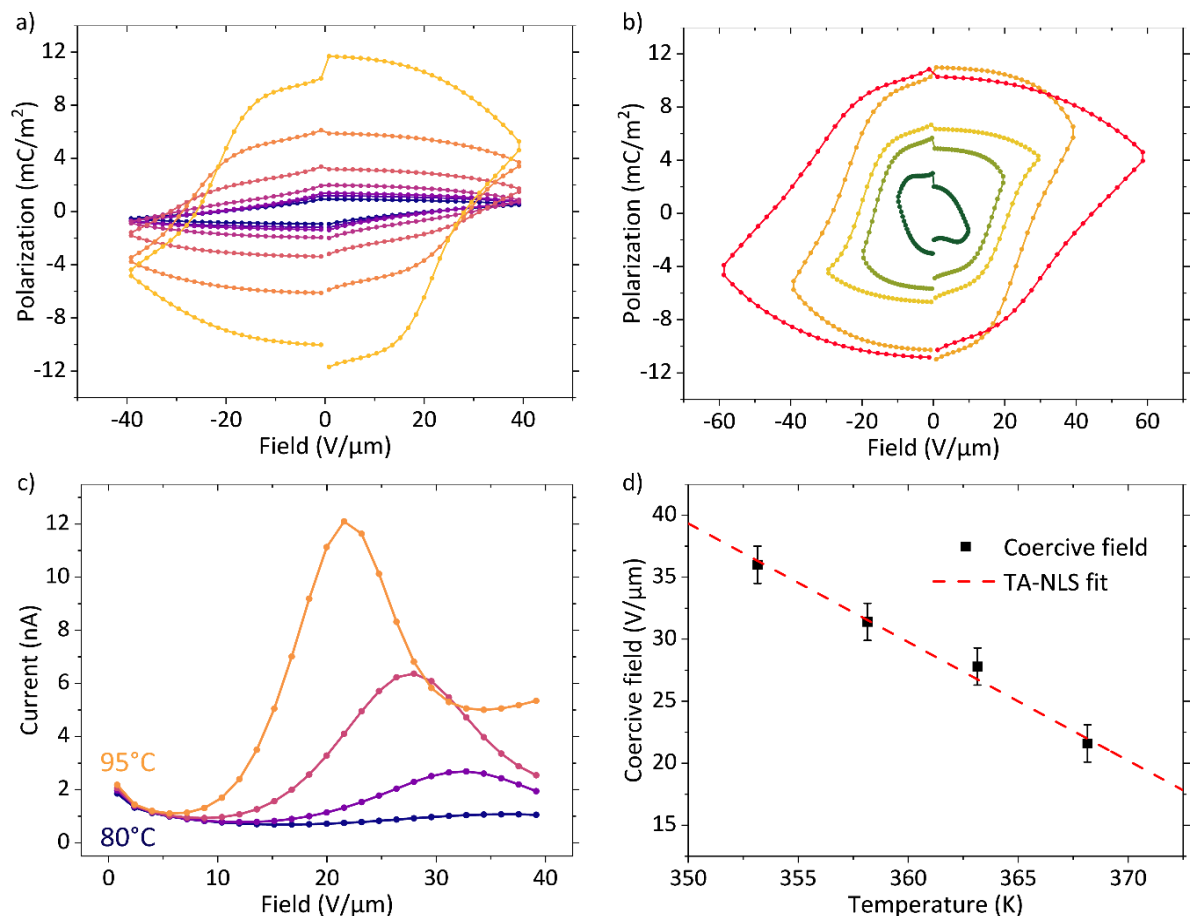

**Figure S9:** a) and b) are the corresponding polarization hysteresis loops to the DWM measurements on **FCH-E** at 20 mHz as shown in Fig. 4a) and b). Although far from ideal, both panels show indications for polarization saturation for increasing electric fields. However, compared to a theoretical polarization of 69 mC/m² for a fully polarized fluorinated cyclohexane ring, the observed polarization values around 5 mC/m² indicate that only a partial reorientation of the fluorine rings occurs. c) shows the current transients obtained from the temperature dependent measurements in Fig. 4a) after DWM correction. The corresponding coercive fields are plotted against the temperature in d) and fitted with the TA-NLS model, taking a typical attempt frequency of 10 THz, the theoretical polarization value of 69 mC/m², and a rise time of $\tau = 6$ s. The resulting fitting parameters are the nucleation volume $V^* = 6.7 \pm 0.3$ nm³ and $w_b = 161 \pm 6$ meV, which are in line with other organic ferroelectrics.[5] Comparing the coercive field values with those at the corresponding temperature for the **FCH-C3-A** material, that is, ~20 K higher due to the higher melting point, reinvigorates the attribution of the high-field peak in Fig. S3 to the pentafluorocyclohexane group.

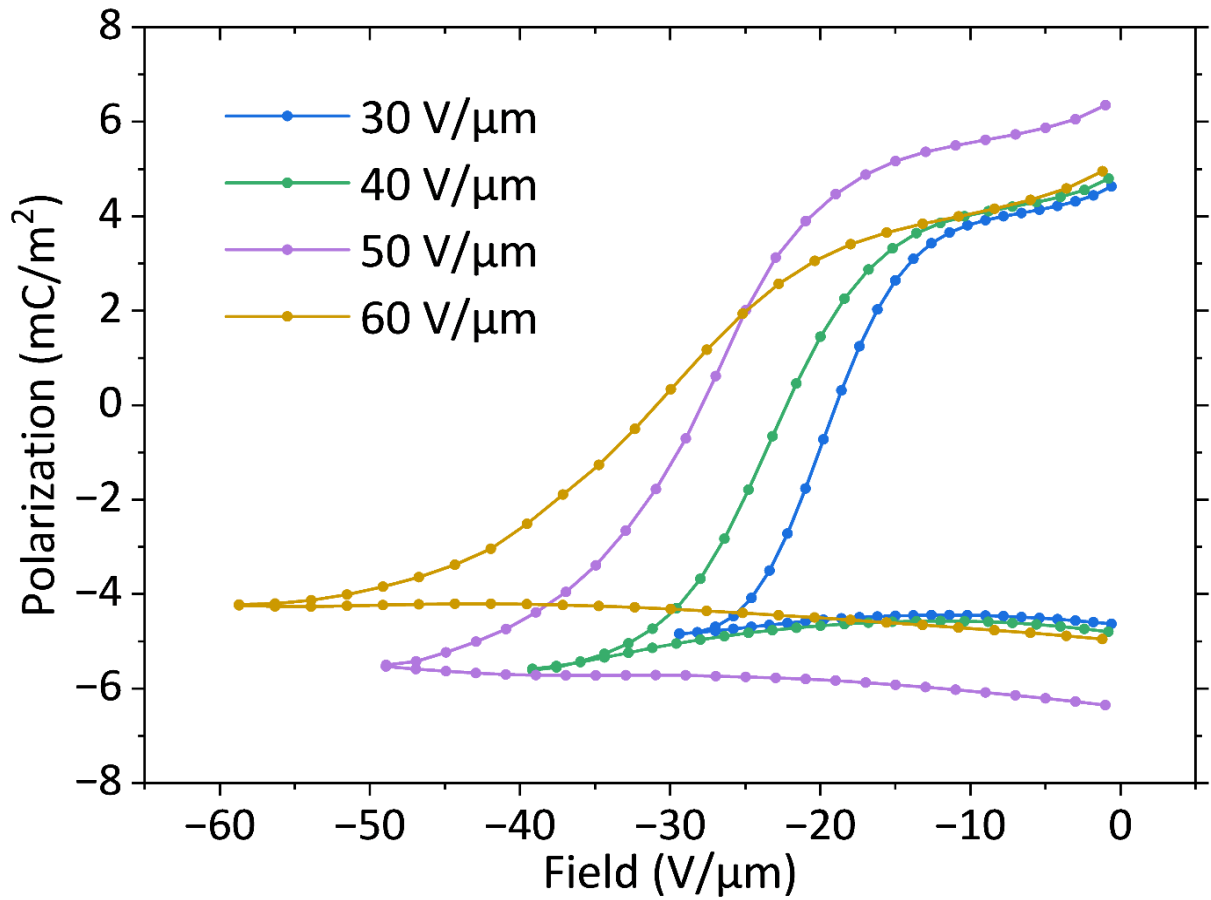

**Figure S10:** Negative part of the polarization hysteresis loop obtained via DWM of **FCH-E** from Fig. 4b) after scaling the background current as schematically shown in Fig. S5a). The resulting polarization is assumed to be the actually measured ferroelectric polarization and amounts to $5.1 \pm 0.8$ mC/m².

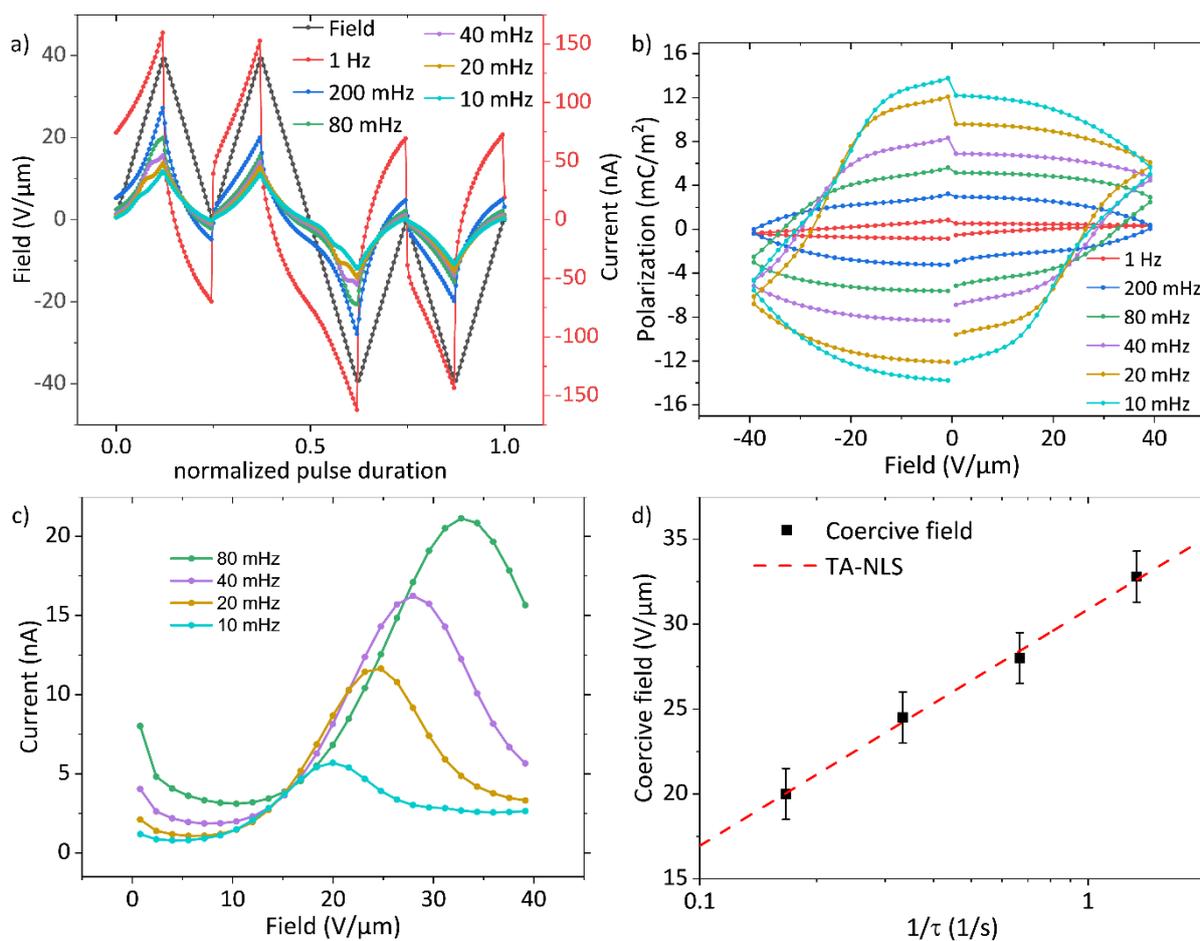

**Figure S11:** a) shows DWM measurements of **FCH-E** at varying frequency at fixed field and a temperature of 95°C with the corresponding hysteresis loops depicted in b). c) shows the current transients obtained after the DWM background correction. The corresponding coercive fields are plotted against the frequency and fit with the TA-NLS model in d). The obtained fitting parameters are an energy barrier $w_b$ of $99.2 \pm 0.3$ meV/nm$^3$ and a nucleation volume $V^*$ of $12 \pm 2$ nm$^3$. These values agree well with those obtained in Fig. S3 for the high field peak of FCH-C3-A, attributed to the fluorinated cyclohexane group. This further reinforces our peak assignment to each dipolar moiety.

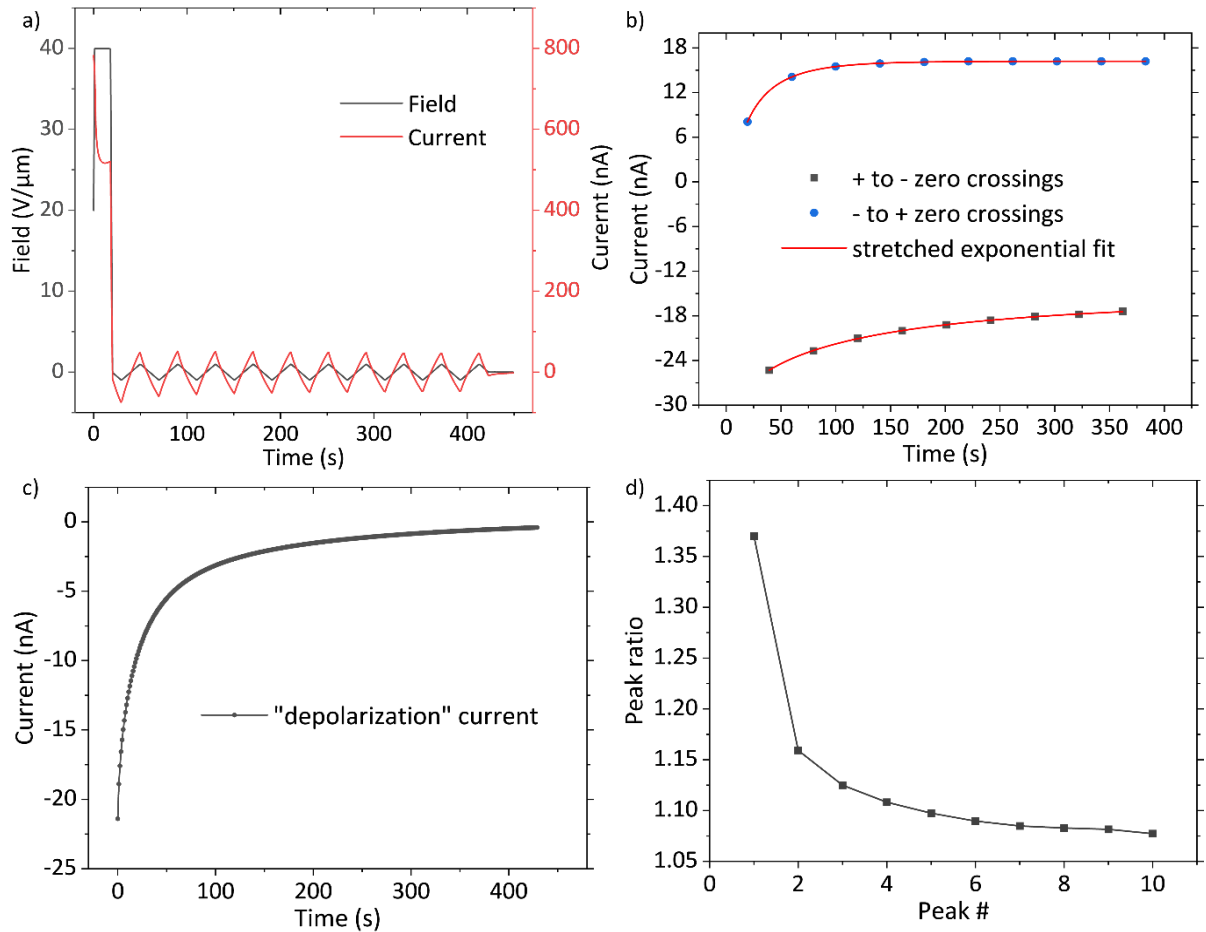

**Figure S12:** Illustration of the steps taken to extract the peak ratios plotted in Fig. 5b) from conductivity measurements on **FCH-C3-A** as depicted in a). First, the currents at the zero crossings of the voltage are obtained and plotted exemplary in b), from positive to negative and vice versa. These are composed of the depolarization current in response to the large rectangular poling pulse, further transient currents like ionic contributions and the displacement current resulting from the field sweep rate. The currents are fitted with a stretched exponential, which is commonly used to describe depolarization processes[7–9], while an added constant term describes the displacement current. With the fit parameters, the depolarization current is reconstructed to the starting time $t = 0$ by averaging the two traces obtained from crossing zero voltage in either direction. This additionally gets rid of the displacement current. An example of a resulting "depolarization" current is plotted in c). The charge obtained by integrating the current transients obtained from multiple measurements averages to around 270 mC/m$^2$ and varies between 125 mC/m$^2$ and 356 mC/m$^2$. As the obtained charge is of the same order but larger than what can solely originate from depolarization (cf. theoretical dipole density of 83 mC/m$^2$), ionic currents or other transients or offset errors are apparently not corrected for. It has to be noted that, as the functional shape is a stretched exponential decay, small variations in starting time result in large differences of total integrated charge, as do any offset errors. Finally, a) is corrected by the transient current background shown in c), and the peak conductivity ratios of the $n^{th}$ peak, that are modulated by the ferroelectric polarization, are shown in d).

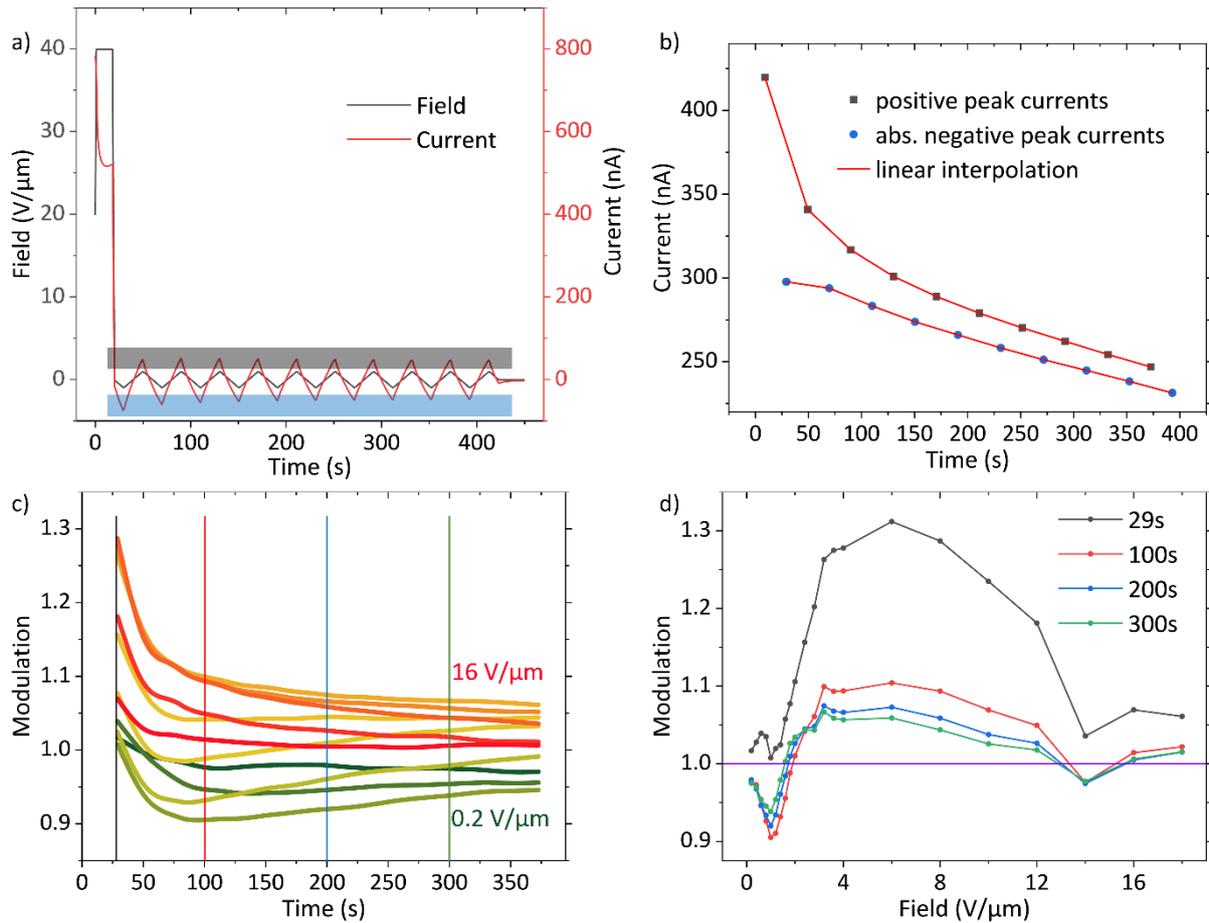

**Figure S13:** Alternative approach to obtain the conductivity modulation in **FCH-C3-A** without the time shift of the up and down peaks. Again, the transient background currents and the displacement current are extracted from the zero crossings of a) as depicted in Fig. S12. The corrected current values of the peaks in a) are plotted in b) and linearly interpolated. From those the current modulation over time shown in c) is obtained for all fields. The modulations were smoothed to account for the linear interpolation edges. The vertical lines indicate the timestamps of 29s, 100s, 200s and 300s for which the field dependence of the modulation is depicted in d), which show an analog behavior as the peak ratios in Fig. 5b).

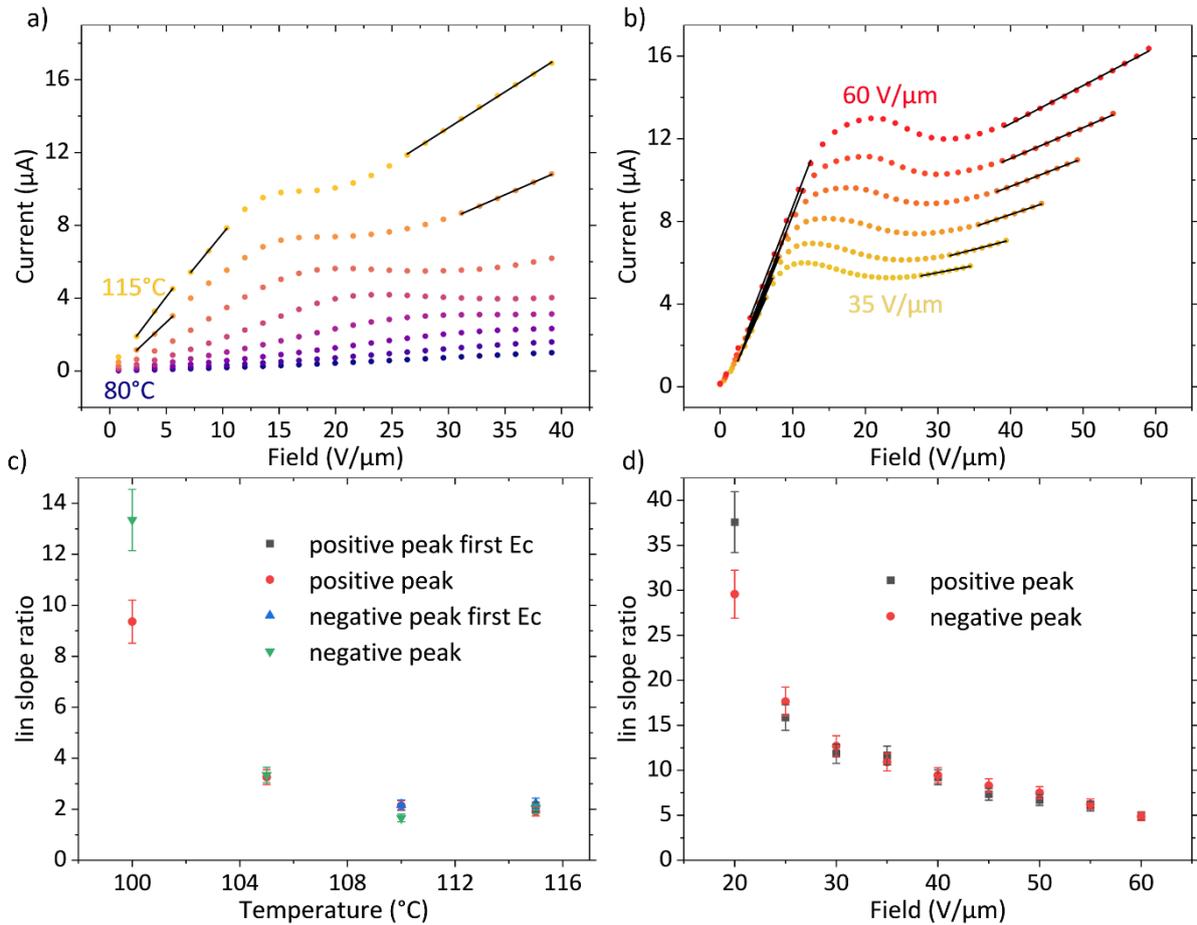

**Figure S14:** On-off ratios obtained from the DWM measurements on **FCH-C3-A** shown in Fig. 2a) and 2c). To that end the left flanks of the switching pulses, i.e. the first positive and first negative (i.e. third overall) pulse, where fitted linearly below and above the coercive field, as depicted in a) for the temperature dependent case (data from Fig. 2a) and in b) for the field dependent case (data from Fig. 2c). In the former, the two different coercive fields have to be kept in mind and are fitted separately, while in the data of the latter only the second, high field coercive field is visible. The ratios obtained by dividing the current slope below the coercive field by the current slope beyond the coercive field are shown in c) and d). Especially for lower fields and temperatures, the switching process might not be completely finalized, resulting in smaller slopes above the coercive field and an inflated on/off-ratio. Error margins are obtained by estimating the variation by changes of the fitting range.

**Supplementary references**